\documentclass[12pt,twoside,fleqn]{p2000}
\usepackage{epsfig}
\usepackage{amssymb}
\usepackage{times}
\mathversion{bold}
\begin{document}

\section*{Mean field studies of high-spin properties in the 
$A\sim 30$ and 60 regions of superdeformation.}

{\large A. V. Afanasjev 
\footnote{on leave of absence from the Laboratory of Radiation Physics,
Institute of Solid State Physics, University of Latvia, LV 2169
Salaspils, Miera str.\ 31, Latvia}, P. Ring,
}\\
{\small
{\it Physik-Department der Technischen Universit{\"a}t
M{\"u}nchen, D-85747 Garching, Germany}\\

{\large I. Ragnarsson}
\\
{\it Department of Mathematical Physics, 
Lund Institute of Technology, S-22100, Lund, Sweden}\\
}

\begin{center}
\begin{minipage}{15cm}
{\bf Abstract:}\\
{\small 
The importance of deformation changes and the possible role 
of proton-neutron pairing correlations on the properties of
paired band crossings at superdeformation in the $A\sim 60$
mass region have been analyzed. The present analysis, supported
in part by the cranked relativistic Hartree-Bogoliubov 
calculations for the SD band in $^{60}$Zn, suggests that when 
going from $^{60}$Zn to neighboring odd nuclei the properties
of paired band crossings are strongly influenced by deformation
changes. A number of questions related to the  superdeformation 
in the $A\sim 30$ mass region has been studied with the cranked 
relativistic mean field theory and the configuration-dependent 
cranked Nilsson-Strutinsky approach.}
\end{minipage}
\end{center}
\vspace*{3mm}

\subsection{Introduction}

Superdeformation at high spin is by now a wide-spread phenomenon
across the periodic table. In recent years, the attention of the
high-spin community has been shifted to the lighter nuclei in 
the vicinity of the $N\approx Z$ line after the discovery of 
superdeformed (SD) bands in $^{62}$Zn \cite{Zn62SD} and neighboring 
nuclei and in $^{36}$Ar \cite{Ar36}. These regions of superdeformation 
are characterized by several distinct features which are either 
not present or much less visible at superdeformation in heavier 
nuclei. First, the relative polarization effects from the individual 
intruder and extruder orbitals are expected to be 
much stronger than in heavier mass regions. Second, the limited
angular momentum content of specific configurations is expected to 
play a significant role in the definition of the rotational 
properties of the SD bands \cite{A60}. For example, a considerable 
decrease of the dynamic moment of inertia ($J^{(2)}$) with respect 
to the kinematic moment of inertia ($J^{(1)}$) has been observed 
in the SD bands of the $A\sim 60$ mass region at the highest
rotational frequencies. This feature is similar to the one seen in 
smoothly terminating bands in the $A\sim 110$ and $A\sim 60$ regions 
\cite{Afa99}. In the $A\sim 30$ mass region, the recently discovered 
SD band in $^{36}$Ar has been observed up to terminating state at 
$I^{\pi}=16^+$ \cite{Ar36}. The third distinct feature is related to 
the possible effects emerging from the proton-neutron pairing 
correlations \cite{Zn61,Cu59}.

 In the present manuscript a number of issues related to
the superdeformation in these mass regions is studied
mainly in the framework of the cranked versions of the
relativistic mean field theory \cite{KR.89,AKR.96,CRHB}.  
In addition, the general structure of the yrast lines
in the $A\sim 30$ mass region emerging from the 
coexistence of collective and non-collective structures
has been analyzed within the configuration-dependent cranked 
Nilsson-Strutinsky approach \cite{Afa99}.
    
\subsection{The $A\sim 60$ $N\approx Z$ mass region.}

 It is well known that the high-spin properties of the nuclei in the 
$A\sim 60$ region are dominated by a variety of phenomena among which smooth 
band termination \cite{Zn62bt,Zn64bt} and superdeformation (SD) 
\cite{Zn62SD,Zn60SD,Zn68} are reliably established today.
The cranked relativistic mean field (CRMF) theory, in which the 
pairing correlations are neglected, has been very successfully 
applied to the description of the properties of a number of SD bands 
in this mass region, see Refs.\ \cite{A60,Cu59,Zn60SD,Zn68} for 
details. In the present contribution,  we will concentrate  
on the properties of SD bands in the $^{60,61}$Zn and 
$^{59}$Cu nuclei which have been in the focus of 
the recent discussions \cite{Zn61,Cu59}.

\begin{figure}
\vspace{-1.0cm}
\begin{picture}(15,8)
\put(0,0){\centerline{\epsfxsize=350pt\epsfbox{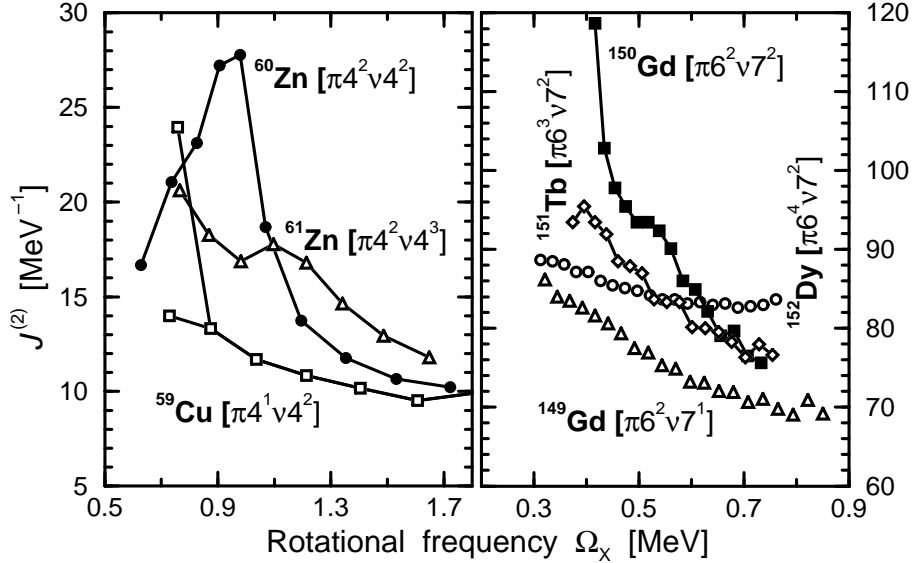}}}
\end{picture}
\vspace{-0.5cm}
\caption{Dynamic moments of inertia of the yrast SD bands in the 
$A\sim 60$ (left panel) and $A\sim 150$ (right panel) mass regions. 
The configurations are labeled by the number of occupied high-$N$ 
intruder orbitals. For details of the configuration assignment in 
the $A\sim 150$ mass region see Refs.\ \protect\cite{AKR.96,ALR.98} 
and references therein.}
\label{a60vsa150}
\end{figure}

The experimental dynamic moments of inertia $J^{(2)}$ of these
bands are shown in Fig.\ \ref{a60vsa150}. One can see that the 
SD band in the $N=Z$\,\, $^{60}$Zn nucleus shows a large jump in
$J^{(2)}$ at rotational frequency $\Omega_x \sim 1.0$ MeV. It was 
suggested in Ref.\ \cite{Zn60SD} that simultaneous alignments 
of the first pairs of the $g_{9/2}$ protons and $g_{9/2}$ neutrons 
are responsible for this observed feature. Such an interpretation was 
based on the fact that in the calculations without pairing the 
assigned configuration has 2 $g_{9/2}$ protons and 2 $g_{9/2}$
neutrons. On the contrary, the SD band in the $N=Z+1$\,\, $^{61}$Zn
nucleus shows only a small bump at $\Omega_x \sim 1.15$ MeV 
indicating a nearly complete blocking of the alignment observed 
in $^{60}$Zn. Because of this feature, the authors of Ref.\ 
\cite{Zn61} questioned the interpretation of the SD band
in $^{60}$Zn given above. Indeed, assuming the same deformation 
of the SD bands in these two nuclei and that the jump in $J^{(2)}$ for the SD 
band in $^{60}$Zn originates from the simultaneous alignments of 
independent $g_{9/2}$ proton and neutron pairs, it is reasonable 
to expect that the odd neutron in $^{61}$Zn should only block the 
neutron contribution to the alignment, while the proton contribution 
should result in an alignment roughly half of that observed in 
$^{60}$Zn. Thus it was suggested that the observed features may be 
due to $T=0$ proton-neutron pairing correlations present in 
$^{60}$Zn, namely the peak in the $J^{(2)}$ of $^{60}$Zn is due to 
the crossing of the $T=1$ and $T=0$ bands. On the other hand, the 
analysis performed in Refs.\ \cite{FS.99a,FS.99b} within the single-$j$
subshell model and the cranked shell model at fixed deformation 
indicates that the frequency of the first band crossing in the $N=Z$ 
nuclei is sensitive to the $T=1$ component of the two-body proton-neutron 
interaction but not to the $T=0$ component. The first alignment 
appears to be delayed by the $T=1$ proton-neutron pairing field
when the intruder shell becomes more symmetrically filled.
Similar to $^{61}$Zn, a SD high-spin band has been observed
experimentally also in the $Z=N-1$\,\, 
$^{59}$Cu nucleus, see Fig.\ \ref{a60vsa150}. However, it is 
more complicated because of a branching at low spin in the observed 
SD band, which gives two $I=25/2$ states, see Fig.\ 1 in Ref.\
\cite{Cu59}. Depending on which state is assumed to be the lowest 
observed state in the SD band, the $J^{(2)}$ moment of inertia is 
either smooth or has a very large jump at the lowest frequencies. In 
the latter case, it might be considered as a possible argument in 
favor of the delay of the first band crossing due to the $T=1$ 
proton-neutron pairing field discussed above.

\begin{figure}[t]
\vspace{1.5cm}
\begin{picture}(15,8)
\put(0,0){\centerline{\epsfxsize=350pt\epsfbox{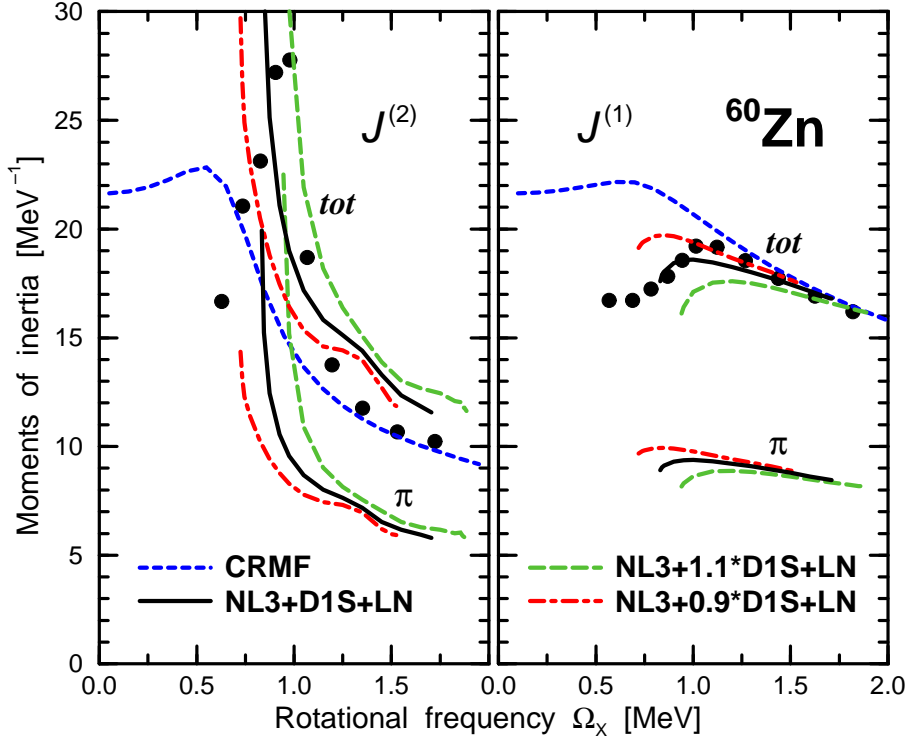}}}
\end{picture}
\vspace{-0.5cm}
\caption{Experimental (solid circles) and calculated (lines) kinematic 
and dynamic moments of inertia of the SD band in $^{60}$Zn. The notation
of the lines is given in the figure. Note that the CRHB results are
shown only in the frequency range where convergence
to the self-consistent SD solution is obtained. At low
rotational frequencies, the solution converges to the
normal-deformed minimum  instead of the SD one, most likely 
due to the small barrier between these two minima.
This, in principle, can be overcomed by an additional constraint 
on the quadrupole moment. At high frequencies, no convergence has 
been obtained after 200 iterations due to very weak pairing 
correlations, see Fig.\ \ref{zn60-anal}. Since the proton and 
neutron contributions to the total dynamic and kinematic moments 
of inertia are almost identical, only the proton contributions 
are shown on the bottom of the panels.}
\label{zn60-j2j1}
\end{figure}

\begin{figure}
\vspace{0.9cm}
\begin{picture}(15,8)
\put(0,-1.0){\centerline{\epsfxsize=380pt\epsfbox{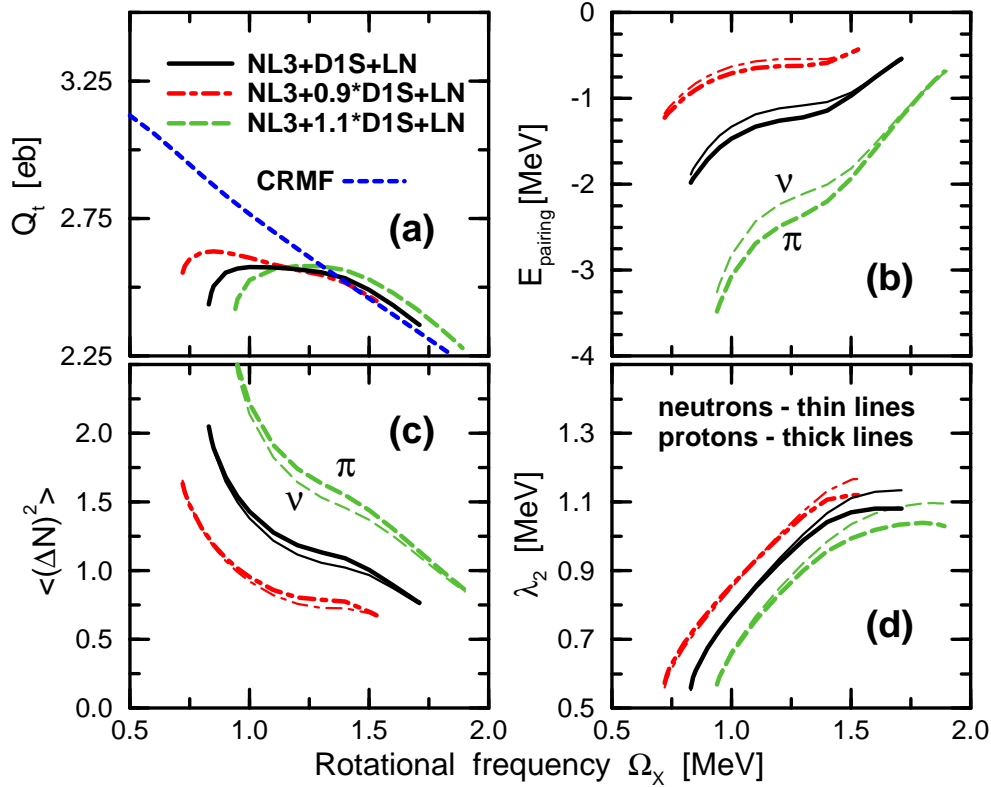}}}
\end{picture}
\vspace{0.4cm}
\caption{Calculated values of the transition quadrupole moments
$Q_t$ (panel (a)), pairing energies 
$E_{pairing}~=~-\frac{1}{2}\mbox{Tr} (\Delta\kappa)$ 
(panel (b)),
particle number fluctuations $\langle (\Delta N)^2 \rangle$
(panel (c)) and $\lambda_2$ (panel (d)) in $^{60}$Zn as a function
of rotational frequency $\Omega_x$ obtained for the solutions
presented in Fig.\ \protect\ref{zn60-j2j1}. The notation of the
lines  is given in the figure.}
\label{zn60-anal}
\end{figure}

The considerations given above neglect the deformation changes 
between the SD bands which can play a considerable role in the 
definition of the properties of the first paired band crossings. 
It is interesting to note that  features similar to the ones
seen in the $^{59}$Cu-$^{60}$Zn-$^{61}$Zn chain have been 
observed earlier in the $A\sim 150$ region of superdeformation,
see Fig.\ \ref{a60vsa150} and Ref.\ \cite{AKR.96} for details.
The large jump and the subsequent small bump in $J^{(2)}$ of the 
yrast SD band in $^{150}$Gd at $\Omega_x < 0.6$ MeV  have been 
explained within the cranked 
Nilsson-Strutinsky approach based on a Woods-Saxon potential 
in terms of consecutive neutron and proton alignments 
\cite{NWJ.89}. The configuration of this band is $\pi 6^2 \nu 7^2$. 
The addition of one $N=6$ proton (band $^{151}$Tb(1)) blocks 
not only the proton paired band crossing (which is expected from simple
blocking arguments) but also the neutron paired band crossing. 
Similarly the configuration of the yrast SD band in $^{152}$Dy does 
not show any signs of the 'expected'  proton and neutron paired band 
crossings. In the same fashion, there are no clear signs of  
proton paired band crossing in the dynamic moment of inertia 
of the $^{149}$Gd(1) band. These examples show the strong 
dependence of the properties of the 'expected' proton and
neutron paired band crossings on the deformation in a system 
where the proton-neutron pairing correlations play no role. 
It is also reasonable to expect a similar impact of deformation 
changes on the properties of SD bands in the $A\sim 60$ region. 
It is especially true considering that the changes in transition 
quadrupole moments $Q_t$ induced by additional 
particle(s) are relatively modest in the $A\sim 150$ region 
($Q_t$($^{152}$Dy)/$Q_t$($^{150}$Gd)$\approx 1.13$,
$Q_t$($^{151}$Tb)/$Q_t$($^{150}$Gd)$\approx 1.06$,
$Q_t$($^{149}$Gd)/$Q_t$($^{150}$Gd)$\approx 0.95$; these ratios are
based on the results of the CRMF calculations, see Ref.\
\cite{AKR.96}) as compared with the $A\sim 60$ region, where, 
for example, the calculated ratio is 
$Q_t$($^{59}$Cu)/$Q_t$($^{60}$Zn)$\approx 0.82$ \cite{Cu59}. 
             
In order to shed some light on the properties of the  
SD band in $^{60}$Zn, they have been studied within the cranked 
relativistic Hartree-Bogoliubov (CRHB) theory 
\cite{CRHB}, which has been very successful in the description 
of SD bands in the $A\sim 190$ mass region
\cite{CRHB,A190}. The calculations have been performed with the NL3 force 
for the RMF Lagrangian and D1S set for the Gogny force in the 
particle-particle channel. Note that only like-particle pairing
has been taken into account. In addition, approximate particle
number projection has been performed by means of the 
Lipkin-Nogami method (further APNP(LN)). In the 
calculations all fermionic and
bosonic states belonging to the shells up to $N_F=12$ and
$N_B=16$ are taken into account and the basis deformation
$\beta_0=0.4, \gamma=0^{\circ}$ is used.
The results of the calculations are shown 
in Figs.\ \ref{zn60-j2j1} and \ref{zn60-anal}.

It is clearly seen in Fig.\ \ref{zn60-j2j1} that the rise in
$J^{(2)}$ in $^{60}$Zn with decreasing rotational frequency and the evolution 
of kinematic moment of inertia $J^{(2)}$ are reasonably well
reproduced in the CRHB calculations with no additional parameters 
(lines indicated by NL3+D1S+LN). At rotational frequencies above 
the band crossing, the CRHB calculations overestimate the experimental 
dynamic moment of inertia by $\sim 10$\%. At these frequencies, the 
CRMF calculations without pairing describe the $J^{(2)}$ and $J^{(1)}$ 
moments almost perfectly. According to the CRHB calculations, the
jump in $J^{(2)}$ originates from the simultaneous alignment of the 
first pairs of the $g_{9/2}$ protons and neutrons. This is contrary 
to the results obtained in the projected shell model \cite{PSMa60} 
where the jump in $J^{(2)}$ originates from two successive band 
crossings.

Fig.\ \ref{zn60-anal} shows the results of the calculations for 
other quantities of interest. The pairing correlations in the $^{60}$Zn 
SD band are much smaller than the ones calculated in the $A\sim 190$
mass region, compare Fig.\ \ref{zn60-anal}b in the present manuscript
with Fig.\ 4 in Ref.\ \cite{CRHB}. They decrease with increasing
rotational frequency reflecting the Coriolis anti-pairing effect.
Note also that the pairing collapse in the proton
and neutron subsystems is observed in the calculations without 
APNP(LN). As a consequence
of weak pairing correlations, the results of the calculations without 
pairing are very close to the ones with pairing at $\Omega_x \geq
1.25$ MeV for the physical observables of interest such as kinematic 
and dynamic moments of inertia (Fig.\ \ref{zn60-j2j1}), and transition
quadrupole moments (Fig.\ \ref{zn60-anal}b). The CRHB calculations 
indicate a slight lowering of kinematic moments of inertia and
transition quadrupole moments $Q_t$ at the frequencies $\Omega_x
<1.25$ MeV as compared with the calculations without pairing. 

The effect of the strength of the Gogny force on the results 
of the CRHB calculations has been investigated by introducing the 
scaling factor $f$ into the Gogny force (see Ref.\ \cite{CRHB}
for details). The results of the calculations with the scaling
factors $f=0.9$ (lines indicated by NL3+0.9*D1S+LN) and
$f=1.1$ (lines indicated by NL3+1.1*D1S+LN) are shown in Figs.
\ref{zn60-j2j1} and \ref{zn60-anal}. The effect of the scaling
of the Gogny force is especially drastic on the pairing properties, 
see Fig.\ \ref{zn60-anal}b,c. It has also considerable impact on 
the rotational properties at $\Omega_x < 1.25$ MeV, see Fig.\ 
\ref{zn60-j2j1}. For example, the frequency of the paired band
crossing is increased (decreased) by $\Delta \Omega_x \sim 0.135$
MeV when the strength of the Gogny force is increased (decreased)
by 10\%. 
 
 The essence of the present discussion of the properties of the 
SD bands in the $^{59}$Cu-$^{60}$Zn-$^{61}$Zn nuclei is the question 
if the observed features are  due to the effects of (i) deformation 
changes or (ii) $T=1$ proton-neutron pairing interaction or 
(iii) combined effect of both of them. The similarity of the
experimental situation in the nuclei around $^{60}$Zn with the
one around $^{150}$Gd strongly suggests that the deformation
changes should play an important role. The main features of
paired band crossing in $^{60}$Zn can be understood in the CRHB 
theory without an explicit proton-neutron pairing. However, this does 
by no means imply that there is no proton-neutron pairing field. 
Further development of the CRHB theory for description of the 
proton-neutron pairing is necessary in
order to clarify how the balance between like-particle
pairing and proton-neutron pairing is changed with increasing
rotational frequency at superdeformation and how it affects 
the observed rotational properties.
 
 One should also note that the approximate particle number projection
by means of the Lipkin-Nogami method might be a poor approximation
to the exact particle number projection in the regime of weak 
pairing correlations in rotating nuclei. For example, the particle 
number fluctuations $\langle (\Delta N)^2 \rangle$ in the unprojected 
wave function are rather small in the $^{60}$Zn SD band indicating
that the breaking of gauge symmetry is small. Studies 
in the full configuration space with exact particle number projection
are definitely needed in order to estimate the accuracy of the Lipkin-Nogami 
method in rotating nuclei in the weak pairing regime.  
  
\subsection{The $A\sim 32$ $N\approx Z$ mass region.}

\begin{figure}[t]
\vspace{0.6cm}
\begin{picture}(15,8)
\put(0,0){\centerline{\epsfxsize=480pt\epsfbox{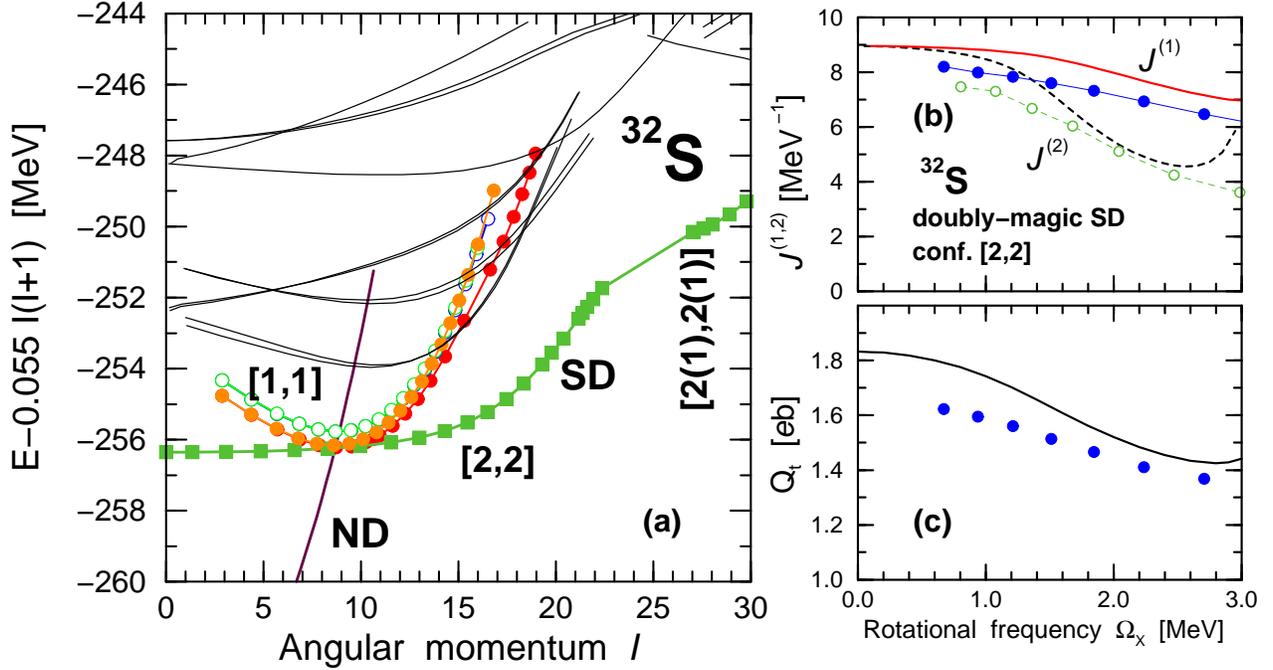}}}
\end{picture}
\vspace{-0.4cm}
\caption{(a) The energies of SD configurations 
calculated in the CRMF theory with the rigid rotor reference
subtracted. In addition, the lowest calculated collective 
states in the normal-deformed (ND) minimum are also shown. 
The configurations discussed in the manuscript are shown
by the lines with symbols. Solid (open) symbols are used for 
total signature $r_{tot}=+1$ ($r_{tot}=-1$). Other excited SD 
configurations are shown by thin solid lines. The configurations 
are labeled by
the shorthand labels $[p_1(p_2),n_1(n_2)]$, where $p_1$ ($n_1$) is 
the number of $f_{7/2}$ protons (neutrons) and $p_2$ ($n_2$) is the
number of $g_{9/2}$ protons (neutrons). Note that $p_2$ ($n_2$)
are omitted when the $g_{9/2}$ orbitals are not occupied.
The kinematic ($J^{(1)}$) and dynamic ($J^{(2)}$) moments
of inertia
and transition quadrupole ($Q_t$) moments  calculated for 
the doubly magic SD configuration [2,2] in $^{32}$S are shown 
in panels (b) and (c), respectively. For comparison, the CNS 
results for these quantities are shown by filled and open
circles.}
\label{S32}
\end{figure}

\begin{figure}[t]
\vspace{0.9cm}
\begin{picture}(15,8)
\put(0,0){\centerline{\epsfxsize=360pt\epsfbox{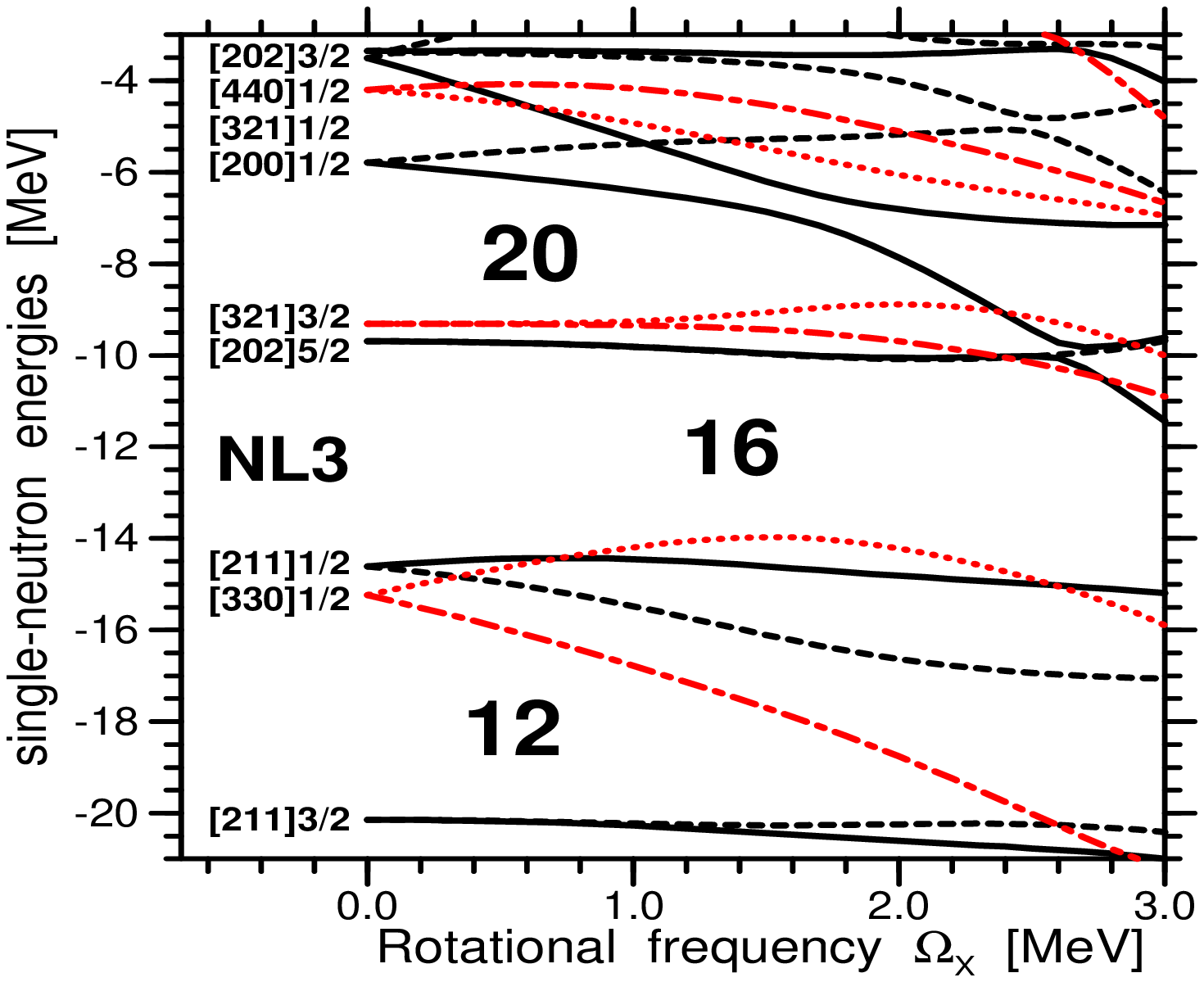}}}
\end{picture}
\vspace{-0.4cm}
\caption{Neutron single-particle energies (routhians) obtained
in the CRMF theory as a function of the rotational 
frequency $\Omega_x$. They are given along the deformation path
of the doubly-magic SD configuration [2,2] in
$^{32}$S. Solid, short-dashed, 
dot-dashed and dotted lines indicate $(\pi=+,\,\,r=-i)$, 
$(\pi=+,\,\,r=+i)$, $(\pi=-,\,\,r=+i)$ and $(\pi=-,\,\,r=-i )$ 
orbitals, respectively.  At $\Omega_x=0.0$ MeV, the single-particle 
orbitals are labeled by means of the asymptotic quantum numbers 
$[Nn_z\Lambda]\Omega$ (Nilsson quantum numbers) of the dominant 
component of the wave function.}
\label{routh-s32}
\end{figure}

In the nuclei in the upper half of the $sd$-shell, high-spin bands 
are formed in configurations with particles excited to the $fp$-shell. 
Of special interest is the yrast SD configuration in $^{32}$S 
which was predicted long ago, see e.g. Refs.\ \cite{She72,LL.74}, but where 
the corresponding rotational band has not been observed at present. 
The question of superdeformation in this nucleus has been 
recently in the focus of a number of theoretical investigations 
within the microscopic theories based on the Skyrme
\cite{MDD.00,YM.99}  and Gogny forces \cite{Gog-s32,Gog-s32-mad}. 
Fig.\ \ref{S32} shows the calculated high-spin structure of this 
nucleus obtained in the CRMF theory with the NL3 parametrization
\cite{NL3} of the RMF Lagrangian. The lowest SD configuration
in this nucleus is lowest in energy amongst collective structures 
in the spin range of $I=10-22\hbar$ and contains two protons and 
two neutrons excited to the $fp$-shell (see Fig. \ref{routh-s32}). 
It has a $\pi 3^2 \nu 3^2$ structure in terms of occupied intruder 
high-$N$ orbitals. The calculated SD configuration is built from 
the magic 16 particle $\omega _{\perp }:\omega _{z}=2:1$ configuration 
of the harmonic oscillator, which appears to be only slightly disturbed in
realistic nuclear potentials. 

The dynamic and kinematic moments of inertia of this configuration are 
shown in Fig.\ \ref{S32}b. While at low rotational frequencies these 
moments are approximately equal, at high  frequencies of 
$\Omega_x \sim 2.5$ MeV, the dynamic moment of inertia is approximately  
60\% of kinematic one. This feature is similar to the one which has been 
observed experimentally in the $A\sim 60$ mass region of superdeformation, 
see Refs.\ \cite{A60,Zn60SD} for details. It comes from the fact
that it is not so much the deformation (at $I=0$) of the configuration
which determines if it is rigid-rotor-like ($J^{(1)}\approx J^{(2)}$)
or not but rather how far away the configuration is from its ``maximum''
spin (see Refs. \cite{A60,R.87} for more details). Indeed, the 
``maximum'' spin of this configuration defined from the distribution 
of particles and holes at low spin is $I_{max}=24\hbar$. The calculations
also indicate a gradual drop of collectivity (i.e., a drop of transition
quadrupole moment $Q_t$) with increasing spin, see Fig.\ \ref{S32}. 
At higher rotational frequencies, the dynamic moments of inertia increases 
due to the admixture of the lowest $g_{9/2}$ orbitals indicating that the 
lowest SD configuration changes its character from [2,2] to [2(1),2(1)]. 


In many details the results of the CRMF calculations for 
the doubly-magic SD configuration in $^{32}$S are similar 
to the ones obtained in the calculations based on the Skyrme 
and Gogny forces \cite{MDD.00,YM.99,Gog-s32}. An interesting
difference is essentially related to the signature splitting 
effects emerging from the time-odd components of the mean 
field. Some features of this effect will be discussed below,
while full results of the investigation of superdeformation in 
the $A\sim 30$ mass region within the framework of the CRMF 
theory will be presented in a forthcoming article \cite{AR.00-a30}.
  
 The signature separation induced by time-odd mean fields has 
been found earlier in the excited SD bands of $^{32}$S in the 
cranked Hartree-Fock calculations based on the effective forces 
of the Skyrme type \cite{MDD.00}. The clear advantage of 
the CRMF theory compared with this approach is the fact that 
time-odd fields (emerging in this theory from nuclear magnetism)
are defined in a unique way. In order to make a comparison 
between these two approaches straightforward, the CRMF calculations 
have been performed for the four excited SD configurations  
having the structure 
\begin{itemize}
\item
$\pi : h[330]1/2^-\,p[202]5/2^{\pm} \otimes \nu :
h[330]1/2^-\,p[202]5/2^{\pm} (r_{tot}=\pm 1)$
\end{itemize}
which display the signature separation induced by time-odd 
fields in Skyrme calculations.
Here the configurations are labeled by the particles $(p)$ and 
holes $(h)$ with respect to the doubly-magic SD configuration 
in $^{32}$S and superscripts to orbital labels are used to 
indicate the sign of the signature $r$ for that orbital 
($r=\pm i$). These configurations are shown by lines with circles
in Fig.\ \ref{S32}. When nuclear magnetism (time-odd mean fields) 
is neglected these four configurations are degenerate in 
energy at no rotation.  This degeneracy is broken and additional 
binding, which depends on the total signature of the configuration 
(0.907 MeV for the $r_{tot}=+1$ configurations and 0.468 MeV for the 
$r_{tot}=-1$ configurations), is obtained when nuclear magnetism is 
taken into account. It is interesting to note that the NL1 and 
NLSH parametrizations of the RMF Lagrangian give very similar
values of additional binding due to nuclear magnetism. 
The essential difference between relativistic and non-relativistic 
calculations lies in the size of the energy gap between the 
$r_{tot}=+1$ and $r_{tot}=-1$ configurations. This gap exceeds 
2 MeV in Skyrme calculations, while it is significantly smaller 
in the CRMF calculations being around 0.45 MeV. In addition, the 
CRMF results for the $r_{tot}=-1$ configurations differ considerably 
from the ones obtained in the Skyrme calculations \cite{MDD.00} 
where it was found that the $r_{tot}=-1$ configurations are not affected 
by the time-odd interactions (i.e. the interactions which give the 
time-odd mean fields through the Hartree-Fock averaging), while 
the $r_{tot}=+i$ configurations are significantly affected and acquire 
an additional binding.

\begin{figure}
\vspace{3.9cm}
\begin{picture}(15,8)
\put(0,0){\centerline{\epsfxsize=470pt\epsfbox{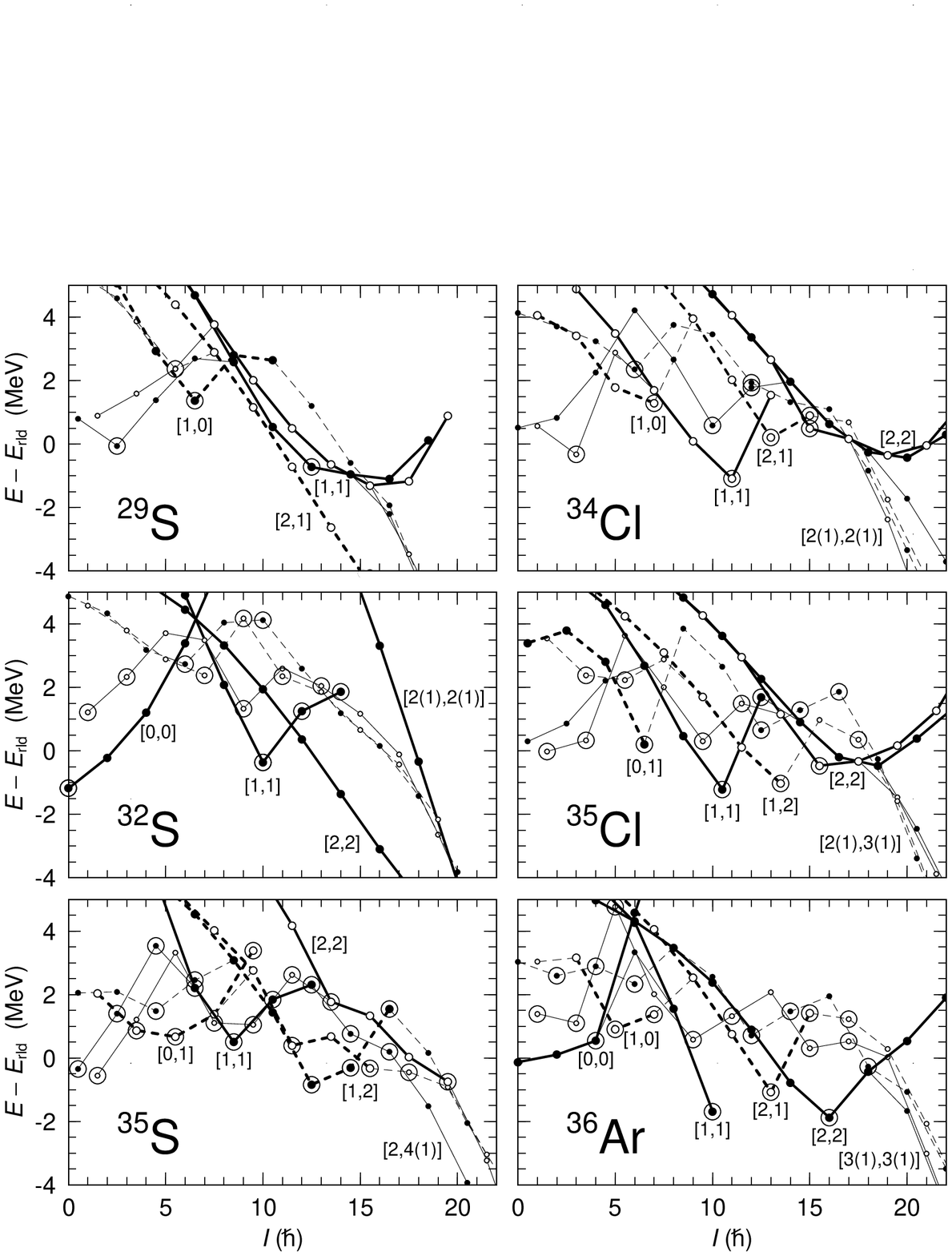}}}
\end{picture}
\vspace{-0.6cm}
\caption{Low-lying configurations for selected nuclei in the 
$A=29-36$ region calculated in the configuration dependent CNS
approach.  The energies are drawn relative to a standard rigid
rotor reference \cite{Afa99}. Note that in Fig.\ \protect\ref{S32}
different reference is used in order to show the SD bands in
more details. Full (dashed) lines are used for positive 
(negative) parity. Filled symbols indicate total signature $\alpha =0$
($r_{tot}=+1$) or $\alpha =1/2$ ($r_{tot}=-i$) and open symbols are
used for total signature $\alpha =1$ ($r_{tot}=-1$) or 
$\alpha =-1/2$ ($r_{tot}=+i$). Aligned non-collective states are 
encircled. The yrast 
lines for the four combinations of signature and parity are drawn by
thin lines while selected fixed configurations are shown by thicker 
lines.  The strongly downsloping 
lines at the highest spin values correspond in general to
configurations at deformations beyond $\varepsilon = 0.6$, typically 
with one or two particles excited to the g$_{9/2}$ subshell. 
}
\label{a30-cns}
\end{figure}

 When discussing the general structure of the high-spectra in 
the $A\sim 30$ mass region, one should remember that similar to 
the $A\sim 60$ mass region it is reasonable to expect that 
non-collective structures will compete with SD ones for yrast 
status in some spin range. Considering that self-consistent
microscopic mean field theories have been applied so far only
to the description of few cases of such coexistence  (see section 
8 in Ref.\ \cite{Afa99} for review) and that such description 
requires further development of relevant computer codes, we use 
here the configuration-dependent cranked Nilsson-Strutinsky 
(CNS) approach for outlining the general structure of high-spin 
spectra in the $A\sim 30$ mass
region. The results of the CNS calculations are shown in Fig.\ 
\ref{a30-cns}.

  In the case of $^{32}$S, the SD configuration with the structure 
[2,2] is yrast from spin $I=12\hbar$. It is interesting to mention that 
in the spin range $I=6-10\hbar$, the yrast line is dominated by 
the aligned single-particle states in the CNS calculations. This 
feature has not been seen in previous microscopic self-consistent 
calculations which were restricted to collective shapes. Both the 
CRMF (Fig.\ \ref{S32}) and CNS calculations show a large gap between 
the doubly magic yrast and the excited SD configurations.  

In heavier nuclei, the observed SD bands are generally formed when the 
orbitals below the 2:1 harmonic oscillator gaps and, in addition, a
few upsloping orbitals above 
these gaps are occupied. For example, in the $N=86$ configuration 
of $^{152}$Dy, the orbitals which are occupied in addition to those 
below the 2:1 $N=80$ gap are [642]5/2, [523]7/2 and [404]9/2.
Similarly, starting from the harmonic oscillator 2:1 gap for 
16 particles, another favored
configuration is formed if the [202]5/2 orbital above this gap is
occupied by two protons and two neutrons.
This configuration appears to be responsible for the recently observed
SD band in $^{36}$Ar \cite{Ar36}. In order to further analyze this
configuration with 4 particles excited to the $fp$-shell, it appears
interesting to investigate how it develops when starting from $^{32}$S, the
[202]5/2 nucleons are added one by one. Thus, the rotational bands for
selected nuclei in this region, calculated in the CNS approach, are 
drawn in Fig.\ \ref{a30-cns}.  Note that in these calculations, the 
proton and neutron single-particle orbitals are almost identical so 
the calculations for one nucleus can also be considered as a prediction 
for its mirror nucleus. 

The configurations in these nuclei can be characterized by the number 
of protons and neutrons in the $fp$-shell which also determines the
number of the holes in the $sd$-shell. It is evident from Fig.\
\ref{a30-cns} that the yrast region is characterized by a number of 
aligned terminating states. These are formed as the maximum spin
states within specific configurations. The orbital occupation in these 
aligned states is straightforward to identify, noting that one proton
or neutron in the $N=3$ shell will at most contribute with $7/2\hbar$ while two
particles of the same kind in this shell contribute with $6\hbar$. These
states are formed at a pretty large oblate deformation $\varepsilon_2 =0.2-0.4$
which means that the two first holes in the $N=2$ shell will be formed in the
[220]1/2 orbital (which is upsloping with increasing oblate
deformation) and therefore do not give any spin contribution in the 
fully aligned state. The next holes are formed in the [211]3/2 and 
[211]1/2 orbitals, thus giving a spin contribution of $3/2\hbar$ for 
the third hole, $2\hbar$ for four holes and $3/2\hbar$ for five
holes. Higher spin states are formed if the last
hole(s) are formed in the [202]3/2 and/or the [202]5/2 orbitals, but the 
states with holes in these orbitals are generally calculated to be less 
favored energetically (because of the large oblate deformation). These
`rules' suggest that favored aligned states should be formed at $I=10\hbar$  
in the [1,1] configurations (with one $f_{7/2}$ proton and one $f_{7/2}$
neutron) both in $^{32}$S (5+5 holes in the $sd$-shell) and in 
$^{36}$Ar (3+3 holes) as confirmed by the full calculations, see Fig.\
\ref{a30-cns}. Comparing the three isotopes $^{29}$S, $^{32}$S and 
$^{35}$S, one can notice a trend that when the number of neutrons 
comes closer to $N=20$, the yrast line is more dominated by aligned 
states while more collective bands are formed for smaller neutron 
numbers.

The SD bands mentioned above are formed in the [2,2] 
configurations. The calculations suggest that such a band is 
relatively low in energy in $^{32}$S, somewhat higher in 
$^{36}$Ar and even higher in the intermediate nuclei $^{35}$Cl, 
$^{34}$Cl and $^{33}$S, the latter nucleus not shown in Fig.\ \ref{a30-cns}. 
The deformation at low spin in these bands varies continuously from 
$\varepsilon_2 \approx 0.4$ in $^{36}$Ar to $\varepsilon_2 \approx 0.6$ 
in $^{32}$S. The $^{36}$Ar band approaches termination
at $I=16\hbar$ continuously, see Ref.\ \cite{Ar36}, while the $^{32}$S
band is calculated only to decrease its deformation marginally 
reaching $\varepsilon_2 \approx 0.57$ ($\gamma \approx 4^{\circ}$)
at $I=18\hbar$ before the deformation starts to increase at even 
higher spin values. The kinematic and dynamic moments of inertia
of this configuration are shown in Fig.\ \ref{S32}b. In $^{34}$Cl, 
four [2,2] bands which are close to signature degenerate are
formed, two of which are shown in Fig.\ \ref{a30-cns}. Their 
deformation decreases from   
$\varepsilon_2 \approx 0.5$ at low spin to $\varepsilon_2 \
\approx 0.4$ ($\gamma = 10-15^{\circ}$) at $I \approx 18\hbar$ and
then increases at  higher spin values. For one of the bands,
see Fig.\ \ref{a30-cns}, this smooth trend is `interrupted' by one 
aligned state 
at $I=15\hbar$ formed according to the rules outlined above. The trends
calculated for these bands are in general agreement with the 
harmonic oscillator rules presented in Ref.\ \cite{Tro79} (see also
Ref.\ \cite{Afa99}) showing a division at a ground state deformation of
$\varepsilon_2 \approx 0.5$ ($\gamma = 0^{\circ}$) between bands which
terminate and bands which stay collective for all spin values, reaching
very large deformations in the limit of very  high spin values.       

\subsection{Conclusions}

 The rotational properties of SD bands in the 
$^{59}$Cu-$^{60}$Zn-$^{61}$Zn nuclei, which are expected to be
influenced by the proton-neutron pairing correlations, have been 
compared with the ones of SD bands around $^{150}$Gd, where the 
proton-neutron pairing plays no role. The similarity of the 
experimental situation in these two regions 
suggests that the presence of a jump in the dynamic moment of
inertia $J^{(2)}$ of the SD band in $^{60}$Zn and
the absence of such jumps in the neighboring nuclei
$^{59}$Cu and $^{61}$Zn cannot be considered as
a clear signal of proton-neutron pairing correlations. It
rather suggests a strong dependence of the 'expected' proton and 
neutron band crossings properties on the deformation of the 
system. Deformation effects should be more pronounced in the $A\sim 60$
mass region where the relative polarization effects induced by the additional 
particle(s) are larger than in the $A\sim 150$ region. For example,
the main features of a paired band crossing in the SD band
of $^{60}$Zn can be understood in the cranked relativistic
Hartree-Bogoliubov theory without an explicit proton-neutron
pairing. This does not imply, however, that there is no 
proton-neutron pairing but suggests that it does not play  
a dominant role in the definition of the properties of paired band 
crossings at superdeformation in the $A\sim 60$ mass region.

  A number of issues such as the high-spin structure 
of $^{32}$S at superdeformation, rotational properties
at superdeformation, the role of nuclear magnetism
(time-odd components of the mean field) in the signature
splitting of   excited SD bands has been investigated
within the cranked relativistic mean field theory. 
In addition, the general structure of high-spin spectra
in the $A\sim 30-35$ mass region has been studied
in the configuration-dependent CNS approach with 
main emphasis on the coexistence of collective 
and non-collective structures along the yrast line
and the particle number dependence of the SD configurations
with two protons and two neutrons in the $fp$-shell.

{\bf Acknowledgements:}\\
{\footnotesize\sf     
A.V.A. acknowledges support from the Alexander von Humboldt
Foundation. This work is also supported in part by the
Bundesministerium f{\"u}r Bildung und Forschung under the
project 06 TM 979 and by the Swedish Natural Science Research
Council.
}


\begin{thebibliography}{99}


\bibitem{Zn62SD} C.\ E.\ Svensson {\it et al.},
Phys.\ Rev.\  Lett. {\bf 79}, 1233 (1997).

\bibitem{Ar36} C.\ E.\ Svensson {\it et al.}, subm. 
to Phys. Rev. Lett. and these proceedings.

\bibitem{A60} A.\ V.\ Afanasjev, I.\ Ragnarsson and P.\ Ring,
Phys.\ Rev. C {\bf 59}, 3166 (1999).

\bibitem{Afa99}
A.\ V.\ Afanasjev, D.\ B.\ Fossan, G.\ J.\ Lane and I.\ Ragnarsson,
Phys.\ Rep.\ {\bf 322}, 1 (1999).

\bibitem{Zn61}
C.-H.\ Yu {\it et al.},  Phys.\ Rev. {\bf C 60}, 031305 
(1999).

\bibitem{Cu59} C.\ Andreoiu {\it et al.},
submitted to Phys. Rev. C. and these proceedings.

\bibitem{KR.89}  W.\ Koepf and P.\ Ring,
Nucl.\ Phys. {\bf A 493}, 61 (1989).

\bibitem{AKR.96} A.\ V.\ Afanasjev, J.\ K{\"o}nig and P.\ Ring,
Nucl.\ Phys. {\bf A 608}, 107 (1996).

\bibitem{CRHB} A.\ V.\ Afanasjev, P.\ Ring and 
J.\ K\"onig, Nucl.\ Phys.\ {\bf A}, in press (see 
also report nucl-th/0001054).

\bibitem{Zn62bt} C.\ E.\ Svensson {\it et al.},
Phys.\ Rev.\ Lett. {\bf 80}, 2558 (1998).

\bibitem{Zn64bt} A.\ Galindo-Uribarri {\it et al.},
Phys.\ Lett. {\bf B 422}, 45 (1998).

\bibitem{Zn60SD} C.\ E.\ Svensson {\it et al.},
Phys.\ Rev.\  Lett. {\bf 82}, 3400 (1999).

\bibitem{Zn68}  M.\ Devlin {\it et al.},
Phys.\ Rev.\ Lett. {\bf 82}, 5217 (1999).

\bibitem{ALR.98} A.\ V.\ Afanasjev, G.\ A.\ Lalazissis and P.\ Ring,
Nucl.\ Phys. {\bf A 634}, 395 (1998).

\bibitem{FS.99a} S.\ G.\ Frauendorf and J.\ A.\ Sheikh,
Nucl.\ Phys. {\bf A 645}, 509 (1999).

\bibitem{FS.99b} S.\ Frauendorf and J.\ A.\ Sheikh,
Phys.\ Rev. {\bf C 59}, 1400 (1999).

\bibitem{NWJ.89} W.\ Nazarewicz, R.\ Wyss and A.\ Johnson,
Nucl.\ Phys. {\bf A 503}, 285 (1989).

\bibitem{A190} A.\ V.\ Afanasjev, J.\ K\"onig and P.\ Ring,
Phys.\ Rev. {\bf C 60}, 051303 (1999).

\bibitem{PSMa60} Y.\ Sun, J.\ Zhang, M.\ Guidry and C.-L.\ Wu,
Phys.\ Rev.\ Lett. {\bf 83}, 686 (1999).

\bibitem{She72} 
R.\ K.\ Sheline, I.\ Ragnarsson and S.\ G.\ Nilsson, 
Phys.\ Lett. {\bf 41B}, 115 (1972).

\bibitem{LL.74} G.\ Leander and S.\ E.\ Larsson,
Nucl.\ Phys. {\bf A 239}, 93 (1975). 

\bibitem{MDD.00} H.\ Molique, J.\ Dobaczewski and J.\ Dudek,
Phys.\ Rev. {\bf C 61}, 044304 (2000) and these proceedings.

\bibitem{YM.99} M.\ Yamagami and K.\ Matsuyanagi,
Nucl.\ Phys. {\bf A 672}, 123 (2000) and these proceedings.

\bibitem{Gog-s32} T.\ Tanaka, R.\ G.\ Nazmitdinov and 
K.\ Iwasawa, report nucl-theor/0004009 (2000). 

\bibitem{Gog-s32-mad} R.\ R.\ Rodriguez-Guzm{\'a}n, J.\ L.\ Egido
and L.\ M.\ Robledo, report nucl-th/0006013 (2000).

\bibitem{NL3} G. A. Lalazissis, J. K\"onig and P. Ring,
Phys. Rev. {\bf C 55}, 540 (1997).

\bibitem{R.87} I.\ Ragnarsson, Phys. Lett. {\bf B 199}, 
317 (1987).

\bibitem{AR.00-a30} A.\ V.\ Afanasjev and P.\ Ring,
in preparation.

\bibitem{Tro79}
T.\ Troudet and R.\ Arvieu, Z.\ Phys. 
{\bf A 291}, 183 (1979); Annals of Physics {\bf 134}, 1 (1981).

\end{thebibliography}
\end{document}